\documentclass{ws-ijmpb}
\usepackage{epsfig}
\usepackage{bm}

\begin{document}

\markboth{Authors' Names}
{Instructions for Typing Manuscripts (Paper's Title)}

%
\catchline{}{}{}{}{}
%

\title{THEORY OF ELECTRON SPECTROSCOPIES\\
IN STRONGLY CORRELATED\\ SEMICONDUCTOR QUANTUM DOTS}

\author{MASSIMO RONTANI}

\address{CNR-INFM National Research Center on nanoStructures
and bioSystems at Surfaces (S3)\\
Via Campi 213/A, Modena, 41100, Italy\\
rontani@unimore.it}

\maketitle

\begin{history}
\received{Day Month Year}
\revised{Day Month Year}
\end{history}

\begin{abstract}
Quantum dots may display fascinating features of strong correlation
such as finite-size Wigner crystallization. We here review a few 
electron spectroscopies 
and predict that both inelastic light scattering and tunneling imaging 
experiments are able to capture clear signatures of crystallization.
\end{abstract}


\section{Quantum Dots as Tunable Correlated Systems}	

Semiconductor quantum dots\cite{Jacak,Tapah} (QDs)
are nanostructures where electrons (or holes) are
confined by electrostatic fields. The confinement field may
be provided by compositional design (e.g., the QD is formed by a
small gap material embedded in a larger-gap matrix) or by gating an
underlying two-dimensional (2D) electron gas. Different techniques
lead to nanometer size QDs with different shapes and strengths of
the confinement. As the typical de Broglie wavelength in
semiconductors is of the order of 10 nm, nanometer confinement leads
to a discrete energy spectrum, with energy splittings ranging from
fractions to several tens of meV. 

The similarity between semiconductor QDs and natural atoms, ensuing from
the discreteness of the energy spectrum, is often pointed 
out.\cite{MaksymPRL,Kastner,Ashoori96,Tarucha96,RontaniNato} Shell
structure,\cite{Ashoori96,Tarucha96,Rontani98} 
correlation effects,\cite{Garcia05} 
Kondo physics,\cite{Goldhaber-Gordon98,Cronenwett98}
are among the most striking experimental demonstrations. 
An intriguing feature of these {\em artificial} atoms is the
possibility of a fine control of a variety of parameters in the
laboratory. The nature of ground and excited few-electron states has
been shown to vary with artificially tunable quantities such as
confinement potential, density, magnetic field, inter-dot 
coupling.\cite{Jacak,Tapah,RontaniNato,ReimannRMP,Rontani04,Ota05}
Such flexibility allows for envisioning a vast range of applications
in optoelectronics (single-electron transistors,\cite{Grabert92} 
lasers,\cite{Kirstaedter94}
micro-heaters and micro-refrigerators based on thermoelectric
effects,\cite{Edwards95}) life
sciences,\cite{Michalet05} as well as in several quantum
information processing schemes in the solid-state environment
(e.g.~Ref.~\refcite{Loss98}).

Almost all QD-based applications rely on (or are influenced by)
electronic correlation effects, which are prominent in these systems.
The dominance of interaction in artificial atoms is
evident from the multitude of strongly correlated few-electron
states measured or predicted under different regimes: Fermi liquid,
Wigner molecule (the precursor of Wigner crystal in 2D bulk), charge
and spin density wave, incompressible state reminescent of
fractional quantum Hall effect in 2D (for reviews see
Refs.~\refcite{ReimannRMP,Rontani01}).
The origin of strong correlation effects in QDs is the following: 
While the kinetic energy term of the Hamiltonian scales as $r_s^{-2}$, 
$r_s$ being the parameter measuring the average distance between electrons, 
the Coulomb energy term scales as $r_s^{-1}$. Contrary to natural atoms, in
QDs the ratio of Coulomb to kinetic energy can be rather
large (even larger than one order of magnitude), the smaller the
carrier density the larger the ratio. This causes Coulomb
correlation to severely mix many different Slater determinants. 

The way of tuning the strength of correlation in QDs we focus on
in this paper is to dilute electron density. At low enough densities,
electrons evolve from a ``liquid'' phase, where low-energy motion
is equally controlled by kinetic and Coulomb energy, to a ``crystallized''
phase, reminescent of the Wigner crystal in the bulk, where
electrons are localized in space and arrange themselves, in absence
of disorder, in a geometrically
ordered configuration (Wigner molecule\cite{ReimannRMP}), so that
electrostatic repulsion is minimized\cite{RontaniNato,ReimannRMP}.  
So far, there are no direct experimental confirmations of the 
existence of these fascinating states.

In this paper we focus on theoretical properties of electron states
in Wigner molecules and on their possible
signatures in experimentally available spectroscopies.
We predict that both inelastic light scattering and 
wave function imaging techniques, like
scanning tunneling spectroscopy (STS), 
may provide direct access
to the peculiar behavior of crystallized electrons.
Specifically, the former is able to probe the ``normal modes'' of the
molecules, while the latter is sensitive to the
spatial order of the electron phase. 

The structure of the paper is as follows.
After an exposition of our theoretical and computational approach
(Sec.~\ref{s:CI}), we briefly review inelastic light scattering
and imaging spectroscopies (Sec.~\ref{s:spectroscopies}),
showing a few results for the four-electron Wigner molecule
(Secs.~\ref{s:raman} and \ref{s:STS}).

\section{Theoretical Approach to the Interacting Problem: Configuration
Interaction}\label{s:CI}

The theoretical understanding of QD electronic states in a vaste
class of devices is based on the envelope function and effective
mass model.\cite{Bastard} Here, changes in the Bloch states,
the eigenfunctions of the bulk semiconductor, brought about by
``external'' potentials other than the perfect crystal potential,
are taken into account by a slowly varying (envelope) function which
multiplies the fast oscillating periodic part of the Bloch states.
This decoupling of fast and slow Fourier components of the wave
function is valid provided the modulation of the external potential
is slow on the scale of the lattice constant. Then, the theory
allows for calculating such envelope functions from an effective
Schr\"odinger equation where only the external potentials appear,
while the unperturbed crystalline potential enters as a renormalized
electron mass, i.e.~the effective value $m^*$ replaces the free
electron mass. Therefore, single-particle (SP) states can be calculated in a
straightforward way once compositional and geometrical parameters
are known. This approach was proved to be remarkably accurate by
spectroscopy experiments for weakly confined QDs;\cite{ReimannRMP,Rontani04}
for strongly confined systems, such as certain classes of
self-assembled QDs, atomistic methods might be
necessary.\cite{Zunger99}

The QD ``external'' confinement potential originates
either from band mismatch or from the self-consistent field 
due to doping charges. In both cases, the total field modulation
which confines a few electrons is smooth and can often be
approximated by a parabolic potential in two dimensions.
Then, the fully interacting effective-mass
Hamiltonian of the QD system reads as:\cite{ReimannRMP}
\begin{equation}
H = \sum_{i}^{N}H_{0}(i)+\frac{1}{2}\sum_{i\neq j}\frac{e^{2}}
{\kappa|\boldsymbol{r_{i}}-\boldsymbol{r_{j}}|},
\label{eq:HI}
\end{equation}
with
\begin{equation}
H_{0}(i)\,=\,\frac{\boldsymbol{p}_i^2}{2m^{*}}
+\frac{1}{2}m^*\omega_0^2r_i^2.
\label{eq:HSP}
\end{equation}
Here, $N$ is the number of free conduction band electrons 
localized in the QD, $e$ and $\kappa$ are respectively the electron charge 
and static relative dielectric constant of the host
semiconductor, $\boldsymbol{r}$ is the position of the electron,
$\boldsymbol{p}$ is its canonically conjugated momentum,
$\omega_0$ is the natural frequency of a 2D harmonic trap.

The eigenstates of the SP Hamiltonian 
(\ref{eq:HSP}) are known as Fock-Darwin (FD) orbitals.\cite{Jacak}
The typical QD lateral extension is given by the characteristic
dot radius $\ell_{\mathrm{QD}} = (\hbar/m^*\omega_0)^{1/2}$,
$\ell_{\mathrm{QD}}$ being the mean square root of $r$ on the 
FD lowest-energy level.
As we keep $N$ fixed and increase $\ell_{\mathrm{QD}}$
(decrease the density), 
the Coulomb-to-kinetic energy dimensionless ratio\cite{Egger99} $\lambda = 
\ell_{\mathrm{QD}}/a^*_{\mathrm{B}}$ [$a^*_{\mathrm{B}}
=\hbar^2\kappa/(m^*e^2)$
is the effective Bohr radius of the dot] increases
as well, driving the system into the Wigner regime.

We solve numerically the few-body problem of Eq.~(\ref{eq:HI}), 
for the ground (or excited) state at
different numbers of electrons, by means of
the configuration interaction (CI) method:\cite{CI}
We expand the many-body wave function in a series of Slater determinants 
built by filling in the FD orbitals with $N$ electrons, and consistently with 
global symmetry constraints. Specifically,  
the global quantum numbers of the few-body system are
the total angular momentum in the direction perpendicular to the plane, $M$,
the total spin, $S$, and its projection along the $z$-axis, $S_z$.
In the Slater-determinant basis the few-body Hamiltonian (\ref{eq:HI}) 
is a large,
block diagonal sparse matrix that we diagonalize by means of a newly
developed parallel code.\cite{donrodrigo}  
Note that, before diagonalization, we are able to build separate 
sectors of the Fock space 
corresponding to different values of $(M,S,S_z)$.
We use, as a single-particle basis, up to 36 FD orbitals, and we are able
to diagonalize matrices of linear dimensions up to $\approx 10^6$
(see Ref.~\refcite{CI}).
The output of calculations consists of both energies and wave functions
of the selected correlated states. We then post-process wave functions 
to obtain those response functions connected to the spectroscopy 
of interest, which we will discuss in Sec.~\ref{s:spectroscopies}. 

\section{Quantum Dot Spectroscopies}\label{s:spectroscopies}

\begin{figure}
\begin{picture}(300,140)
\put(70,0){\epsfig{file=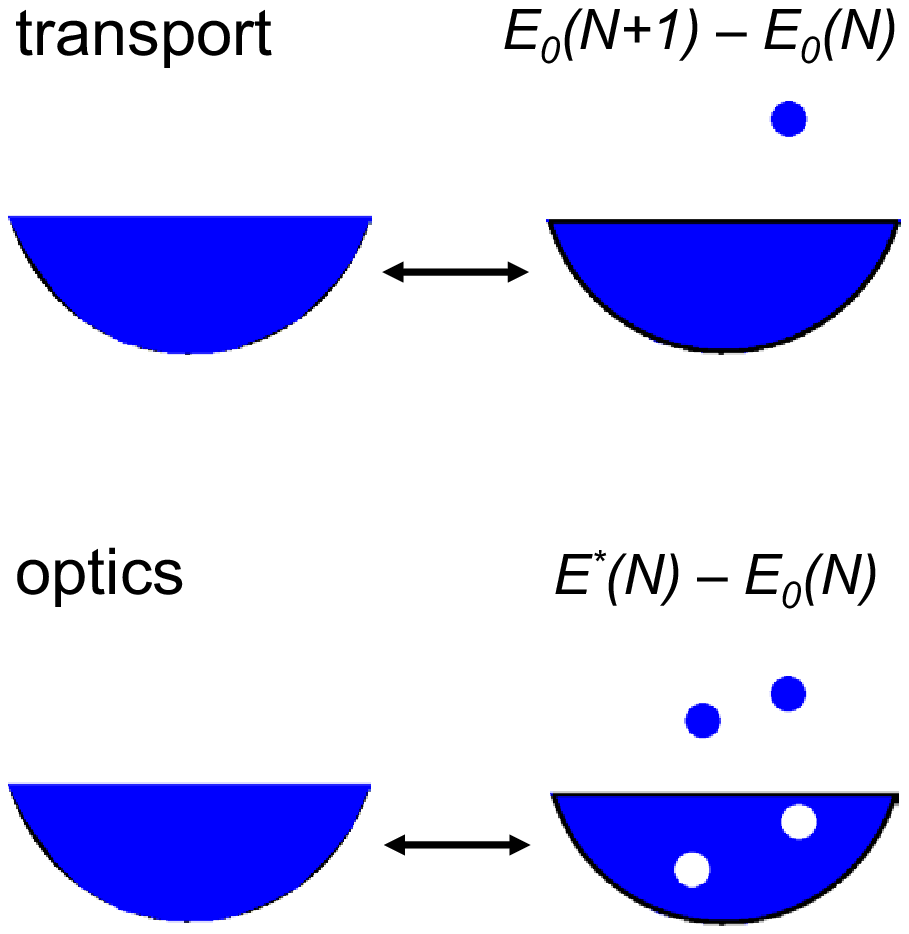,width=1.8in,,angle=0}}
\end{picture}
\caption{Transport vs. optical electron spectroscopies in 
quantum dots.}
\label{f:spectra}
\end{figure}

There are two main classes of electron spectroscopies in QDs 
(cf.~Fig.~\ref{f:spectra}). In single-electron tunneling 
spectroscopies
one is able to inject just one electron into the interacting system,
in virtue of the Coulomb blockade effect.\cite{Grabert92} 
In this way one accesses quantities related mainly to 
the ground states of both $N$ and $N+1$ electrons, like the dot chemical
potential $E_0(N+1)-E_0(N)$, where $E_0(N)$ is the $N$-electron 
ground state energy.
While such transport spectroscopies were very successful in QD studies,
they suffer severe limitations when probing excited states.
On the other hand, optical techniques such as far-infrared and
inelastic light scattering spectroscopies, which leave the probed system
uncharged, allow for an easier access to excitation energies $E^*(N)$.  
Below we focus on two specific examples of these families of 
experiments (for reviews see Refs.~\refcite{Jacak,Tapah,ReimannRMP,Rontani01}).  
\subsection{Inelastic light scattering}

Inelastic light scattering experiments in QDs probe low-lying 
neutral excitations. Depending on the relative orientation
of the polarizations of the incoming and scattered photons, one
is able to access either charge or spin density fluctuations.\cite{Abstreiter}
Different excited states can be probed by varying the frequency
of the laser with respect to the optical energy band gap of the host
semiconductor, or by changing the momentum tranferred from
photons to electrons.\cite{Garcia05,Lockwood,Schuller,Hamburg}  

Only recently QDs with very few electrons were studied.\cite{Garcia05,Hamburg}
Here we focus on charge density excitations probed when the
laser frequency is far from the resonance with the optical 
gap.\cite{Hawrylak,EPLus}
In this limit, the scattering cross section $\mathrm{d}\sigma
/\mathrm{d}\omega$ at a given energy $\omega$ is proportional to
the momentum-resolved dynamical response function:
\begin{equation}
\frac{\mathrm{d}\sigma}{\mathrm{d}\omega} \approx
\sum_n\left|M_{n0}\right|^2\delta\left(\omega-\omega_n+\omega_0\right)$,
with $M_{n0}=\sum_{ab\sigma} \left<a|{\rm e}^{{\rm
i}\mathbf{q}
\cdot\mathbf{r}}|b\right>\left<n|c^{\dagger}_{a\sigma}c_{b\sigma}
|0\right>.
\label{eq:raman}
\end{equation} 
Here $\left|0\right>$ and $\left|n\right>$ are the ground and
excited interacting few-body states, as obtained by the CI 
computation, with energies $\omega_0$ and $\omega_n$, respectively,
$c^{\dagger}_{a\sigma}$ is a fermionic operator
creating an electron in the $a$-th FD orbital with spin $\sigma$,
$\mathbf{q}$ is the wave vector transferred in the inelastic
photon scattering event. Formula (\ref{eq:raman}) describes how
ground and excited states are selectively coupled by 
charge density fluctuations of momentum $\mathbf{q}$.

\subsection{Wave function imaging via tunneling spectroscopy}

The imaging experiments, in their essence, measure
quantities directly proportional to the
probability for transfer of an electron through a barrier, from an
emitter, where electrons fill in a Fermi sea, to a dot, with
completely discrete energy spectrum.
In multi-terminal setups one can neglect the role of electrodes
other than the emitter, to a first approximation.
The measured quantity can be
the current,\cite{Grandidier,Vdovin} the differential
conductance,\cite{Tarucha96,Millo,Maltezopoulos} or the QD
capacitance,\cite{Ashoori96,Wibbelhoff,Maan} while the emitter can be the STS
tip,\cite{Grandidier,Millo,Maltezopoulos} or a $n$-doped GaAs
contact,\cite{Ashoori96,Tarucha96,Vdovin,Wibbelhoff,Maan}
and the barrier can be the
vacuum\cite{Grandidier,Millo,Maltezopoulos} as well as a AlGaAs
spacer.\cite{Ashoori96,Tarucha96,Vdovin,Wibbelhoff,Maan}

The measured quantities are generally understood as proportional
to the density of carrier states
at the resonant tunneling (Fermi) energy,
resolved in either real\cite{Grandidier,Millo,Maltezopoulos} or
reciprocal\cite{Vdovin,Wibbelhoff} space.
However, Coulomb blockade phenomena and strong
inter-carrier correlation ---the fingerprints of
QD physics--- complicate the above simple picture. 
Below we review the theoretical framework we have
recently developed\cite{brief,jjap} to clarify the quantity actually
probed by imaging spectroscopies.

According to the seminal paper by Bardeen,\cite{Bardeen}
the transition probability (at zero temperature) is given by the expression
$(2\pi/\hbar)\left|{\cal{M}}\right|^2n(\epsilon_f)$, where
$\cal{M}$ is the matrix element and $n(\epsilon_f)$ is the energy
density of the final QD states. The standard theory would predict the
probability to be proportional to the total density of QD states
at the resonant tunneling energy, $\epsilon_f$, possibly space-resolved
since $\cal{M}$ would depend on the resonant QD orbital.\cite{Tersoff}
Let us now assume that: (i) Electrons in the emitter do not interact
and their energy levels form a continuum.
(ii) Electrons from the emitter access through the barrier a single QD at 
a sharp resonant energy, corresponding to a well defined interacting QD state.
(iii) The QD is quasi-2D, the electron motion being separable in the $xy$ 
plane and $z$ axis, which is
parallel to the tunneling direction. (iv) Electrons
in the QD all occupy the same confined orbital along $z$, 
$\chi_{\mathrm{QD}}(z)$.
Then one can show\cite{brief} that the matrix element $\cal{M}$ may
be factorized as
\begin{equation}
{\cal{M}}\propto T M,
\end{equation}
where $T$ is a purely single-particle matrix element while
the integral $M$ contains the whole correlation physics.

The former term is proportional to the current density evaluated
at any point $z_{\mathrm{bar}}$ in the barrier:
\begin{equation}
T = \frac{\hbar^2}{2m^*}\left[\chi^*_{\mathrm{E}}(z)
\frac{ \partial \chi_{\mathrm{QD}}(z) }{\partial z}
-\chi_{\mathrm{QD}}(z)
\frac{\partial \chi_{\mathrm{E}}^*(z) }{\partial z}
\right]_{ z=z_{\mathrm{bar}} },
\label{eq:T}
\end{equation}
where $\chi_{\mathrm{E}}(z)$ is the resonating emitter state along $z$ 
evanescent in the barrier.
The term (\ref{eq:T}) contains the information regarding the
overlap between emitter and QD orbital tails in the barrier,
$\chi_{\mathrm{E}}(z)$ and $\chi_{\mathrm{QD}}(z)$, respectively.
Since $T$ is substantially independent from both $N$ and $xy$ location,
its value is irrelevant in the present context. 

On the other hand, the in-plane matrix element $M$ conveys 
the information related to correlation effects.
If we now specialize to STS and assume an ideal, point-like tip, then 
$M\propto \varphi_{\mathrm{QD}}(\bm{r})$ where 
$\varphi_{\mathrm{QD}}(\bm{r})$ is the quasi-particle (QP)
wave function of the interacting QD system:
\begin{equation}
\varphi_{\mathrm{QD}}(\bm{r}) = \langle N - 1 | \hat{\Psi}(\bm{r})|
N \rangle .
\label{eq:def}
\end{equation}
Here 
$\hat{\Psi}(\bm{r})$
is the fermionic field operator destroying an electron
at position $\bm{r}\equiv (x,y)$, $| N - 1 \rangle $
and $| N \rangle $ are the QD interacting ground states
with $N-1$ and $N$ electrons, respectively, calculated via CI
(Sec.~\ref{s:CI}). We omit spin indices for the sake of
simplicity.

Therefore, the STS differential conductance is proportional
to $\left|\varphi_{\mathrm{QD}}(\bm{r})\right|^2$,
which is the usual result of the standard non-interacting
(or mean-field) theory\cite{Maltezopoulos,Tersoff} for the highest-energy
occupied (or lowest-energy unoccupied) orbital, provided the SP 
(mean-field) orbital is replaced by the QP wave
function unambiguously defined by Eq.~(\ref{eq:def}).

\section{Inelastic Light Scattering Spectroscopy in the Wigner 
Limit}\label{s:raman}

\begin{figure}
\centerline{\epsfig{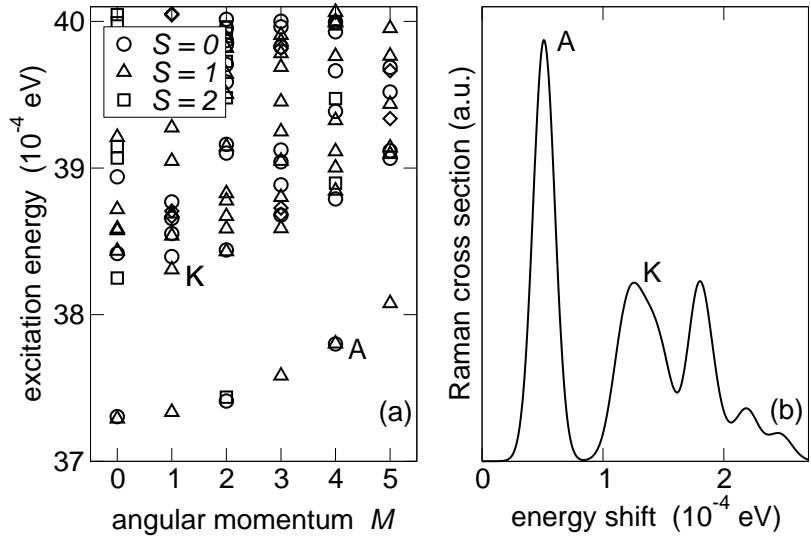}}
\caption{(a) Excitation energies of the four-electron Wigner molecule
vs. total orbital angular momentum $M$, for different values 
of the total spin $S$, with $\lambda=10$. (b) Corresponding Raman 
scattering cross section
for charge excitations in the off-resonance limit. 
Here we set $\left|\mathbf{q}\right|=2\times 10^5$ cm$^{-1}$ 
and introduce a fictitious gaussian broadening of Raman 
peaks ($\sigma=0.9\times 10^{-5}$ eV).
GaAs parameters were used throughout the paper.
}
\label{f:raman}
\end{figure}

We focus on the four-electron system at density low enough ($\lambda=10$)
to induce the crystallization of a  Wigner molecule.
Figure~\ref{f:raman}(a) displays our CI results for the low-energy
region of the excitation spectrum 
for different values of the total orbital angular momentum
and spin multiplicities (the so called {\em yrast} line\cite{ReimannRMP}).
The ground state is the 
spin triplet state with $M=0$, which is found to be the 
lowest-energy state in the range $0\le \lambda \le 20$.
The absence of any level crossing as the density is progressively
diluted (i.e.~$\lambda$ increases) implies that the crystallization
process evolves in a continuos manner, consistently with the finite-size
character of the system. Nevertheless, several features of 
Fig.~\ref{f:raman}(a) demonstrate the formation of a Wigner molecule
in the dot. 

First, the two possible spin multiplets 
other than $S=1$, namely $S=0,2$,
lie very close in energy to the ground state. 
This is understood by invoking
electron crystallization. In fact, in the Wigner limit, the Hamiltonian
of the system turns into a classical quantity, since the 
kinetic energy term in it may be
neglected with respect to the Coulomb term. Therefore, 
only commuting operators (the electron positions) appear
in the Hamiltonian. In this regime the spin,
which has no classical counterpart, becomes 
irrelevant:\cite{CI,Rontani05a} spin-dependent
energies show a tendency to degeneracy. This can also be understood 
in the following way: if electrons sit at some lattice sites with
unsubstatial overlap of their localized wave functions, then the
total energy must not depend on the relative orientation of
neighboring spins.

Second, an exact replica of the sequence of the two lowest-energy
singlet and triplet states occurs for both $M=0$ and
for $M=4$ [(cf. the triplet state labeled A in Fig.~\ref{f:raman}(a)]. 
Such a period of four units on the $M$ axis identifies
a magic number,\cite{MaksymPRL,Koskinen01} whose origin is brought about
by the internal spatial symmetry of 
the interacting wave function.\cite{Ruan95,MaksymReview} 
In fact, when electrons form
a stable Wigner molecule, they arrange themselves into a
{\em four-fold} symmetry configuration, where charges are localized 
at the corners of a square.\cite{EPLus,Bolton93,Bedanov94}

A third distinctive signature of crystallization is the appearance
of a {\em rotational band},\cite{Koskinen01} 
which in Fig.~\ref{f:raman}(a) is composed of
those lowest-energy levels, separated by energy gaps of about 
$0.1$ meV, that increase monotonically as $M$ increases. 
This band is  called ``rotational'' since it
can be identified with the quantized levels $E_{\mathrm{rot}}(M)$ 
of a rigid two-dimensional top, given by the formula
\begin{displaymath}
E_{\mathrm{rot}}(M)=\frac{\hbar^2}{2I}M^2,
\end{displaymath}
where $I$ is the moment of inertia of the top.\cite{Koskinen01} These
excitations may be thought of as the ``normal modes'' of
the Wigner molecule rotating as a whole in the $xy$ plane
around the vertical symmetry axis parallel to $z$.

Figure \ref{f:raman}(b) displays the calculated off-resonance Raman spectrum
for charge density fluctuations ($\Delta S = 0$, where $\Delta S$ is the
variation of the spin with respect to the ground state).
The dominant peak is the one labeled A, corresponding to the $\Delta M = 4$
normal mode of the Wigner molecule of Fig.~\ref{f:raman}(a), which 
therefore represents a clear, visible feature of Wigner crystallization.
Notice also the appearance of the so called Kohn mode\cite{Jacak} (labeled K) at
higher energy, which is a dipolar ($\Delta M = 1$) collective motion
of the center of mass of the electron system.

\section{Wigner Molecule Formation Seen by Imaging
Spectroscopy}\label{s:STS}

\begin{figure} 
\begin{picture}(300,173)
\put(60,0){\epsfig{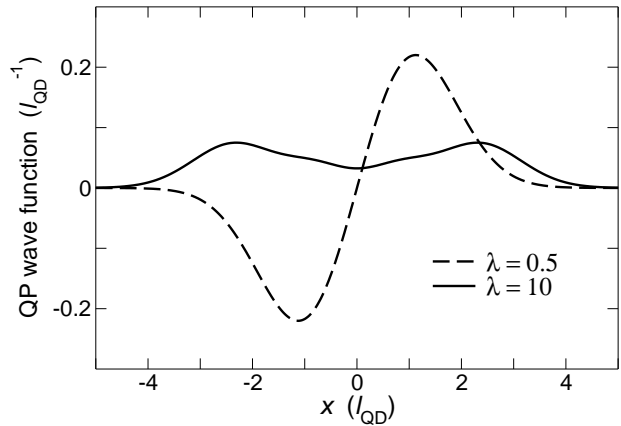}}
\end{picture}
\caption{Quasi-particle wave function vs. $x$ ($y=0$) for different
values of the dimensionless Coulomb-to-kinetic energy ratio $\lambda$.
The STS differential conductance, proportional to the wave function square modulus,
corresponds to the tunneling process $N=3\rightarrow N=4$.
The length unit is the characteristic dot radius $\ell_{\mathrm{QD}}$.
}
\label{f:qp}
\end{figure}

Figure \ref{f:qp} shows the QP wave function, 
corresponding to the injection of the fourth electron into the QD, 
as a function of $x$ (at $y=0$),
for two different values of $\lambda$.
As $\lambda$ increases (from 0.5 to 10), the density decreases
going from the non-interacting
limit ($\lambda=0.5$)
deep into the Wigner regime ($\lambda=10$). 
At high density ($\lambda=0.5$, approximately corresponding
to the electron density $n_e= 3.3 \times 10^{12}$ cm$^{-2}$) 
the wave function substantially
coincides with the non-interacting FD $p$-like orbital.
By increasing the QD radius (and $\lambda$), the QP wave function
weight moves towards larger values of $r$ (the $s$-like symmetry of the
$\lambda=10$ curve comes from a transition of the three-electron
ground state\cite{CI} around $\lambda \approx 6$). 
By measuring lengths in units of $\ell_{\mathrm{QD}}$, as it is done in 
Fig.~\ref{f:qp}, this trivial effect should be totally 
compensated. However, we see for $\lambda=10$
($n_e \approx 1.1 \times 10^{9}$
cm$^{-2}$) that the now much stronger
correlation is responsible for an unexpected weight reorganization,
which is related to the formation of a ``ring'' of crystallized
electrons in the Wigner molecule (cf.~Sec.~\ref{s:raman}).
The shape of the $\lambda=10$ curve of Fig.~\ref{f:qp} is consistent
with the onset of a solid phase with four electrons sitting at the
apices of a square, as discussed in Sec.~\ref{s:raman}.
Note also the dramatic weight loss of QPs as $\lambda$ is increased:
the stronger the correlation, the more effective the ``orthogonality''
between interacting states. 

\section{Conclusions}

The Wigner molecule is an intriguing electron phase, peculiar of
QDs, which still lacks for experimental confirmation.
We predict that both inelastic light scattering and imaging
spectroscopies are able to probe distinctive features of crystallization.

\section*{Acknowledgements}

G. Goldoni and E. Molinari in Modena have been deeply involved in 
the elaboration of the results presented here. We thank for fruitful 
discussions V. Pellegrini, C. P. Garc\'{\i}a, A. Pinczuk, S. Heun, 
A. Lorke, G. Maruccio, S. Tarucha, B. Wunsch.
This paper is supported by Supercomputing Project 2006 (Iniziativa 
Trasversale INFM per il Calcolo Parallelo), MIUR-FIRB Italia-Israel
RBIN04EY74, Italian Ministry of Foreign Affairs
(Ministero degli Affari Esteri, DGPCC).

\end{document}